\documentclass[a4paper,11pt]{article}
\usepackage{pos}
\DeclareMathOperator{\diag}{diag}
\newcommand{\SU}{\mathrm{SU}}
\newcommand{\alphas}{\alpha_{\mbox{\rm s}}}
\newcommand{\tin}{t_{\mbox{\tiny{in}}}}
\newcommand{\tfin}{t_{\mbox{\tiny{fin}}}}
\newcommand{\Scl}{S_{\mbox{\tiny{cl}}}}

\title{Calculation of the running coupling in non-Abelian gauge theories from Jarzynski's equality}
\ShortTitle{Running coupling from Jarzynski's equality}

\author[a]{Olmo Francesconi}
\author*[b,c]{Marco Panero}
\author[c]{David Preti}

\affiliation[a]{CP3-Origins and Department of Mathematics and Computer Science, University of Southern Denmark,\\
Campusvej 55, 5230 Odense M, Denmark}

\affiliation[b]{Department of Physics, University of Turin,\\
Via Pietro Giuria 1, I-20125 Turin, Italy}

\affiliation[c]{INFN, Turin,\\
Via Pietro Giuria 1, I-20125 Turin, Italy}

\emailAdd{francesconi@imada.sdu.dk}
\emailAdd{marco.panero@unito.it}
\emailAdd{david.preti@to.infn.it}

\abstract{We discuss the theoretical foundations of non-equilibrium Monte Carlo simulations based on Jarzynski's equality and present, as an example of application, the determination of the running coupling in the Schr\"odinger-functional scheme.
}

\FullConference{%
 The 38th International Symposium on Lattice Field Theory, LATTICE2021
  26th-30th July, 2021
  Zoom/Gather@Massachusetts Institute of Technology
}

\begin{document}
\maketitle

\section{Introduction and motivation}
\label{sec:introduction_and_motivation}

In this contribution we discuss our recent work~\cite{Francesconi:2020fgi}, presenting an application of a new type of non-equilibrium Monte~Carlo algorithms for lattice field theory. The formalism underlying our work rests on some theorems that were formulated in non-equilibrium statistical mechanics during the 1990's~\cite{Evans:1993po, Evans:1993em, Gallavotti:1994de, Gallavotti:1995de, Jarzynski:1996oqb, Jarzynski:1997ef} which, in addition to giving new mathematical insights into irreversibility and non-equilibrium phenomena, also provide the basis for a novel type of Monte~Carlo calculations.

Specifically, our work is based on an algorithm, derived from a theorem by C.~Jarzynski~\cite{Jarzynski:1996oqb, Jarzynski:1997ef}, that has already found various numerical applications, including for the evaluation of interface tensions in spin systems~\cite{Chatelain:2007ts} and, more recently, the computation of the equation of state in lattice gauge theory~\cite{Caselle:2016wsw, Caselle:2018kap}. Here, we present another application: the determination of the running coupling in $\SU(N)$ Yang-Mills theories in the Schr\"odinger-functional scheme~\cite{Symanzik:1981wd, Luscher:1992an}.

\section{Jarzynski's theorem}
\label{sec:Jarzynski_s_theorem}

Consider a statistical system whose Hamiltonian $H$ depends on a set of parameters $\lambda$. Jarzynski's theorem states that, when the system is driven out of equilibrium by making the parameters of the Hamiltonian time-dependent according to a given protocol $\lambda(t)$ (for $\tin \le t \le \tfin$), the exponential average of the work $W$ done on the system in units of the temperature $T$ equals the ratio of the partition functions for \emph{equilibrium} states with parameters $\lambda(\tfin)$ and $\lambda(\tin)$:
\begin{equation}
\label{Jarzynski_theorem}
\overline{\exp \left( -W/T \right)} = \frac{Z_{\lambda(\tfin)}}{Z_{\lambda(\tin)}} .
\end{equation}

Before presenting a proof of this equality, it is worthwhile pointing out some remarks about it. First of all, a surprising aspect of eq.~(\ref{Jarzynski_theorem}) is that it is an equality between equilibrium quantities (the partition functions on the right-hand side) and genuinely non-equilibrium ones (the average on the left-hand side is taken over the non-equilibrium trajectories that the system can follow). The time $t$ mentioned in the statement of Jarzynski's theorem can either be real time, like in experimental verifications of this equality~\cite{Liphardt:2002ei}, or Monte~Carlo time in a simulation, which is the case we will focus on. The average denoted by the bar on the left-hand side is taken over all of the possible trajectories that the system can follow, when its parameters are modified according to the $\lambda(t)$ protocol, but the latter is \emph{arbitrary} and fixed once and for all, i.e. there is no averaging over the possible choices for the $\lambda(t)$ function. Combining eq.~(\ref{Jarzynski_theorem}) with Jensen's inequality~\cite{Jensen:1906sl}, one immediately obtains the second law of thermodynamics for the non-equilibrium evolution of the system. Finally, Jarzynski's equality reduces to known results in two different limits. When $\tfin-\tin$ tends to infinity, the system remains arbitrarily close to thermodynamic equilibrium throughout every trajectory: in that case, the work done on the system is always exactly equal to the free-energy difference between the initial (for $\lambda=\lambda(\tin)$) and final ($\lambda=\lambda(\tin)$) ensembles, $W=\Delta F$, and eq.~(\ref{Jarzynski_theorem}) becomes trivial. Conversely, when the parameters are instantaneously switched from $\lambda(\tin)$ to $\lambda(\tfin)$ (i.e. for $\tfin-\tin \to 0$), Jarzynski's equality reduces to the formula expressing statistical reweighting~\cite{Ferrenberg:1988yz}. Indeed, the numerical Monte~Carlo algorithm based on Jarzynski's equality that we are discussing can be thought of as \emph{a non-equilibrium generalization of reweighting}.

Among the various proofs of eq.~(\ref{Jarzynski_theorem}) that have been formulated, here we present a sketch of one that is ``constructive'' for our purposes, in the sense that it gives an idea of how eq.~(\ref{Jarzynski_theorem}) can be implemented in a non-equilibrium Monte~Carlo calculation on the lattice. To set the notation, we define the standard Boltzmann distribution for a field configuration $\phi$ as $\pi[\phi] = \exp \left ( - H[\phi] / T \right) / Z$ and the normalized transition probability from $\phi$ to a configuration $\phi^\prime$ as $P[\phi\to\phi^\prime]$. Note that both $\pi$ and $P$ depend on $\lambda$ and, as a consequence, in our setup are functions of time. We assume that the following, detailed-balance-like, condition holds:
\begin{equation}
\label{detailed_balance}
\pi[\phi] P[\phi\to\phi^\prime] = \pi[\phi^\prime] P[\phi^\prime \to \phi] .
\end{equation}
The exponential appearing on the left-hand side of eq.~(\ref{Jarzynski_theorem}) can be written by discretizing the $[\tin,\tfin]$ interval into $N$ sub-intervals (with $t_0=\tin$ and $t_N=\tfin$), and taking the $N \to \infty$ limit:
\begin{equation}
\exp \! \left( -W/T \right) \! = \! \lim_{N \to \infty} \! \exp \! \left( \! - \! \sum_{n=0}^{N-1} \frac{H_{\lambda\left(t_{n+1}\right)}\! \! \left[\phi\left(t_n\right)\right]-H_{\lambda\left(t_n \right)}\! \! \left[\phi\left(t_n \right)\right]}{T} \right) \! =\!  \lim_{N \to \infty} \!  \prod_{n=0}^{N-1} \! \frac{Z_{\lambda(t_{n+1})} \pi_{\lambda(t_{n+1})}\! \! \left[\phi\left(t_n \right)\right]}{Z_{\lambda(t_n)} \pi_{\lambda(t_n)}\! \! \left[\phi\left(t_n \right)\right]}.
\end{equation}
The average over non-equilibrium trajectories can then be expressed by the (integral) sum over all of the possible field configurations $\phi(t_n)$ at all times $\tin \le t_n \le \tfin$, that we denote as $\sum_{\left\{ \phi(t) \right\} }$:
\begin{equation}
\label{expression_of_exponential_average_I}
\overline{\exp \! \left( -W/T \right)} \! = \! \lim_{N \to \infty} \! \! \sum_{\left\{ \phi(t) \right\} } \pi_{\lambda(\tin)}\! \! \left[ \phi(\tin) \right] \prod_{n=0}^{N-1} \left\{ \frac{Z_{\lambda(t_{n+1})}}{Z_{\lambda(t_n)}} \! \cdot \! \frac{\pi_{\lambda(t_{n+1})}\! \! \left[\phi\left(t_n \right)\right]}{\pi_{\lambda(t_n)}\! \! \left[\phi\left(t_n \right)\right]} \! \cdot \! P_{\lambda(t_{n+1})}\! \! \left[ \phi(t_n) \to \phi(t_{n+1}) \right] \right\} ,
\end{equation}
where the field configuration at the initial time $\tin$ is assumed to be in thermal equilibrium, but those at all times $t>\tin$ are not. Then, using eq.~(\ref{detailed_balance}) and simple algebraic manipulations to simplify the telescoping products of ratios on the right-hand side of eq.~(\ref{expression_of_exponential_average_I}), one obtains
\begin{equation}
\label{expression_of_exponential_average_II}
\overline{\exp \left( -W/T \right)} = \frac{Z_{\lambda(\tfin)}}{Z_{\lambda(\tin)}}\lim_{N \to \infty}\sum_{\left\{ \phi(t) \right\} } \pi_{\lambda(\tfin)}\left[ \phi(\tfin) \right] \prod_{n=0}^{N-1} P_{\lambda(t_{n+1})}\left[ \phi(t_{n+1}) \to \phi(t_n) \right] .
\end{equation}
The sums on the right-hand side of eq.~(\ref{expression_of_exponential_average_II}) can be carried out explicitly, starting from the one on $\phi(t_0)$, which appears only in the $P_{\lambda(t_1)}\left[ \phi(t_1) \to \phi(t_0) \right]$ term, using the fact that the transition probabilities are normalized, and proceeding iteratively to sum over the field configurations at all times $\tin < t < \tfin$. Finally, the last (integral) sum over $\phi(t_N)$ is a sum of the $\pi_{\lambda(\tfin)}\left[ \phi(\tfin) \right]$ Boltzmann factors, which are normalized to unity, too, so that eq.~(\ref{expression_of_exponential_average_II}) reduces to eq.~(\ref{Jarzynski_theorem}).

Following this proof of Jarzynski's theorem, one can construct a Monte~Carlo algorithm to compute the right-hand side of eq.~(\ref{Jarzynski_theorem}) by evaluating the exponential average of $-W/T$ over a sufficiently large number of non-equilibrium trajectories, in which the field configurations, initially in thermal equilibrium, are progressively driven out of equilibrium by varying the parameters of the Hamiltonian as a function of time, without allowing the system thermalize. Through the usual map between statistical mechanics and Euclidean field theory (whereby, for instance, $H/T$ is replaced by the Euclidean action $S$, etc.), this computational strategy can be directly applied in numerical calculations in lattice field theory, too.

\section{Running coupling in the Schr\"odinger-functional scheme}

\begin{figure}[t]
\centerline{\includegraphics[height=0.25\textheight]{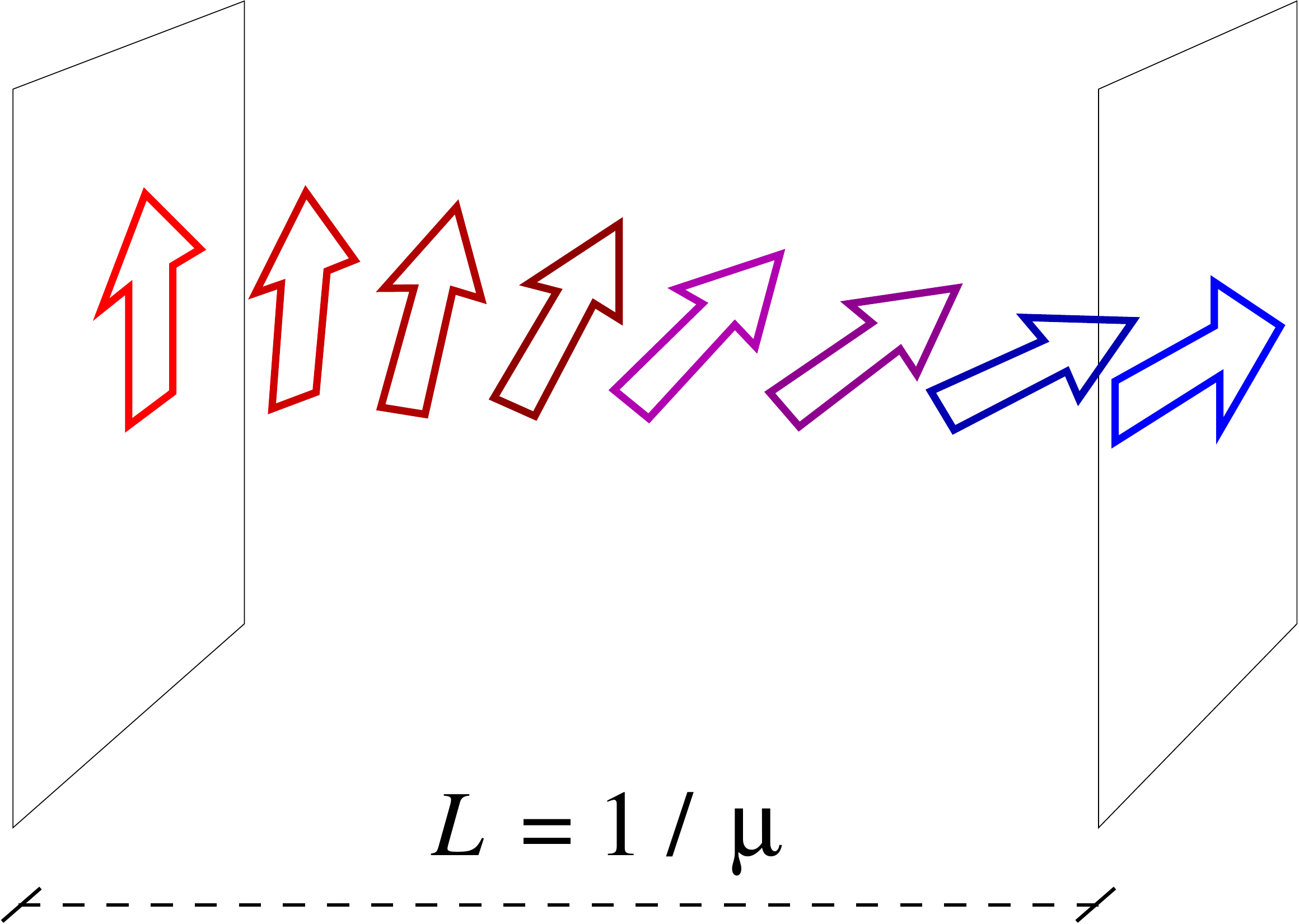}}
\caption{In the Schr\"odinger-functional scheme, fixed boundary conditions are imposed on the gauge fields at the initial and at final Euclidean time, and a running coupling is defined from the quantum effective action of the system.}
\label{fig:SF_classical_field}
\end{figure}

Consider the functional integral $Z$ describing the Euclidean-time evolution of a gauge theory, with bare coupling $g_0$, between initial ($\tau=0$) and final ($\tau=L$) Euclidean times, at which spatially uniform Dirichlet boundary conditions are imposed on the gauge fields (while periodic boundary conditions are assumed along the spatial directions, whose extent is also taken to be $L$). Classically, this induces a field configuration, sketched in fig.~\ref{fig:SF_classical_field}, with Euclidean action $\Scl$ that is proportional to $1/g_0^2$. In the Schr\"odinger-functional scheme~\cite{Symanzik:1981wd, Luscher:1992an} one defines the running coupling $g$ at the length scale $L$ by assuming its inverse square to be proportional to the corresponding quantum effective action $\Gamma$ (and fixing its normalization in such a way that, at the lowest order in a perturbative expansion of $\Gamma$ at weak coupling, $g$ reduces to $g_0$).

We computed $g^2(L)$ with Jarzynski's equality, through Monte~Carlo simulations in which the system is driven out of equilibrium by changing the boundary conditions at $\tau=0$ and at $\tau=L$, and monitoring the corresponding response in the effective action $\Gamma = -\ln Z$. Apart from the different way to evaluate the coupling, our setup is the same as the one that was used in the computation of the running coupling in the Schr\"odinger-functional scheme by conventional Monte~Carlo calculations in refs.~\cite{Luscher:1992zx, Luscher:1993gh}. For the $\SU(2)$ Yang-Mills theory the field configurations for spatial link matrices at the initial and final Euclidean times are set to $U=\exp(iaC_{\tau})$, with 
\begin{equation}
\label{boundary_conditions_in_terms_of_eta}
C_0 = \frac{1}{L} \diag \left( -\eta , \eta \right),
\quad
C_L = \frac{1}{L} \diag \left( \eta-\pi , \pi-\eta \right)
\end{equation}
and the parameter $\eta$ (which plays the r\^ole of $\lambda$) is varied from $\eta(\tin)=\pi/4$ to $\eta(\tfin)=\eta(\tin)+\Delta \eta$ through a sequence of ``quenches'' (i.e. jumps) in Monte~Carlo time, with $\Delta \eta$ ranging from $10^{-3}$ to $1.5 \cdot 10^{-2}$. Accordingly, the squared running coupling $g^2(L)$ is obtained from the formula
\begin{equation}
\label{SU2_SF_coupling}
g^2(L) = - \lim_{\Delta \eta \to 0} \frac{24 \Delta \eta}{\Delta \Gamma} \left( \frac{L}{a} \right)^2 \sin \left[ \frac{\pi}{2} \left( \frac{a}{L} \right)^2 \right] .
\end{equation}

Similarly, for the $\SU(3)$ theory the boundary matrices $U=\exp(iaC_{\tau})$ are defined through
\begin{equation}
\label{boundary_conditions_in_terms_of_eta_and_nu}
C_0 = \frac{1}{L} \diag \left(
\eta - \frac{\pi}{3},
\eta \nu - \frac{\eta}{2},
\frac{\pi}{3} - \eta \nu - \frac{\eta}{2}
\right), 
\quad
C_L = \frac{1}{L} \diag \left(
-\eta - \pi,
\frac{\pi}{3} + \eta \nu + \frac{\eta}{2},
\frac{2\pi}{3} - \eta \nu + \frac{\eta}{2}
\right)
\end{equation}
and the running coupling is obtained as
\begin{equation}
\label{SU3_SF_coupling}
g^2(L) =  \lim_{\Delta \eta \to 0} \frac{12 \Delta \eta}{\Delta \Gamma} \left( \frac{L}{a} \right)^2 \left\{ \sin \left[ \frac{2\pi}{3} \left( \frac{a}{L} \right)^2 \right] + \sin \left[ \frac{\pi}{3} \left( \frac{a}{L} \right)^2 \right] \right\}
\end{equation}
by varying $\eta$ from $\eta(\tin)=0$ to $\eta(\tfin)=\eta(\tin)+\Delta \eta$ (keeping $\nu=0$ fixed).

The evolution of the coupling is then determined through a step-scaling function $\sigma (s, g^2(L))=g^2(sL)$~\cite{Luscher:1991wu} (an integrated version of the $\beta$ function of the theory), which in principle allows one to follow the running of this coupling over a wide interval of length scales.

\section{Results}

\begin{figure}[t]
\centerline{\includegraphics[height=0.25\textheight]{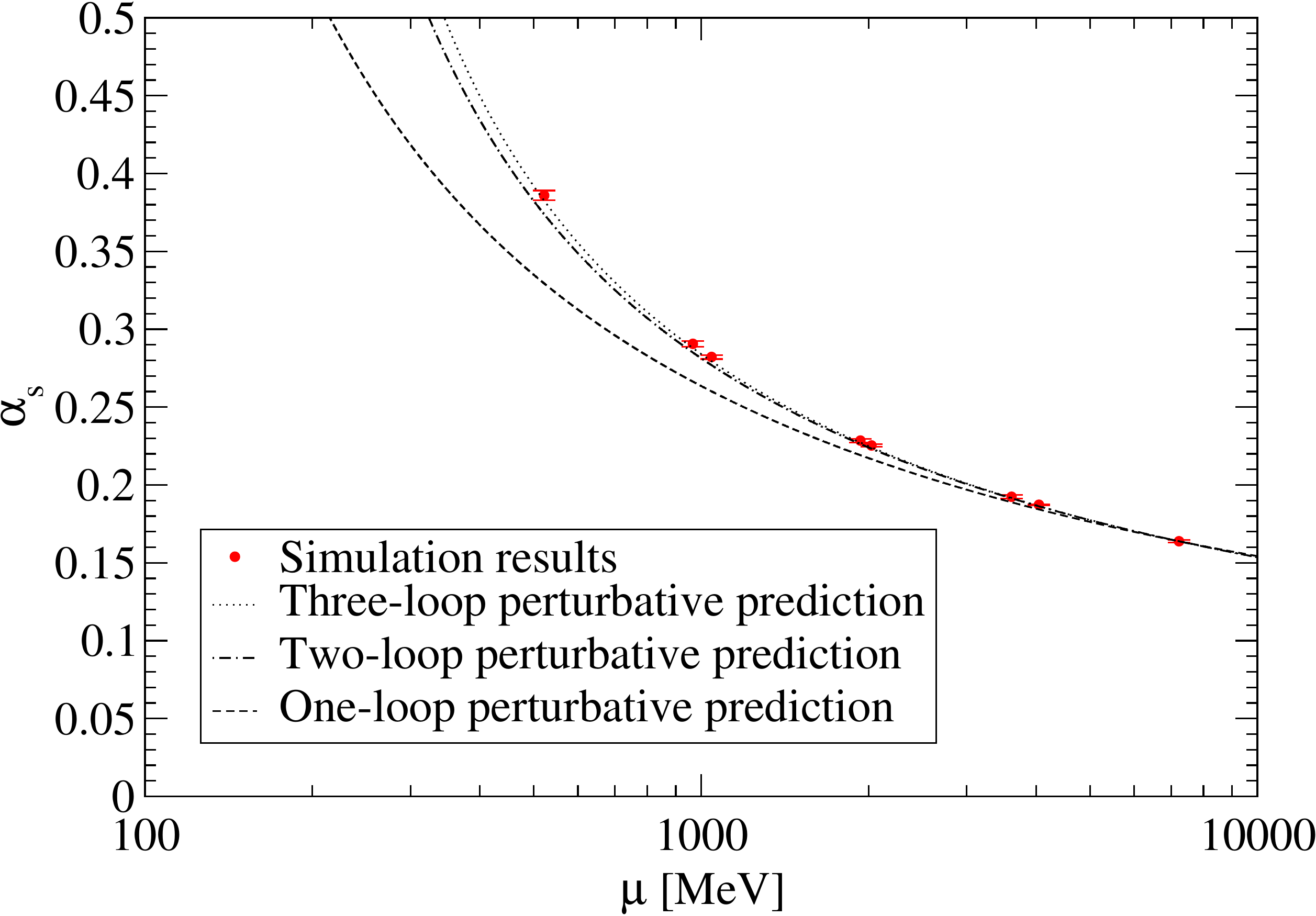} \hfill \includegraphics[height=0.25\textheight]{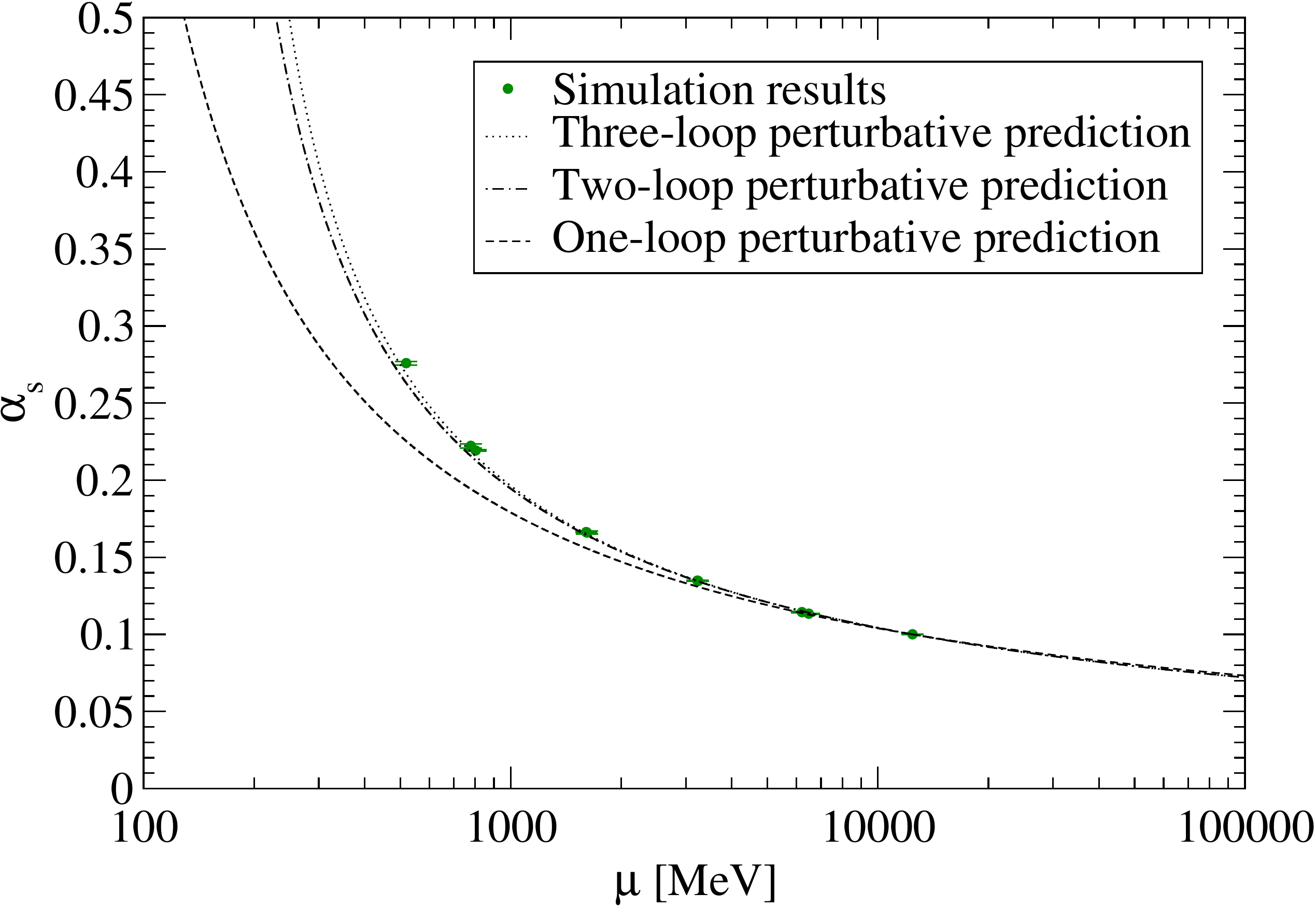}}
\caption{Left-hand-side panel: Results for $\alphas=g^2/(4\pi)$ in the Schr\"odinger-functional scheme (red dots) for $\SU(2)$ Yang-Mills theory and perturbative predictions at one (dashed black curve), two (dash-dotted black line), and three loops (dotted black curve), according to eq.~(\ref{integrated_beta_function}). Right-hand-side panel: Numerical results for $\SU(3)$ Yang-Mills theory (green symbols) and one-, two-, and three-loop predictions (denoted by the same types of lines as in the plot on the left-hand side) derived from eq.~(\ref{three-loop_SU3_SF_beta_function}).}
\label{fig:results}
\end{figure}

Our results for $\alphas=g^2/(4\pi)$ in $\SU(2)$ Yang-Mills theory, which are fully compatible with those reported in ref.~\cite{Luscher:1992zx}, are shown in the plot on the left-hand-side panel of fig.~\ref{fig:results}, where they are denoted by red symbols, plotted as a function of $\mu=1/L$, and compared with the three-loop perturbative prediction (and its truncations at two loops and at one loop), which can be written as:
\begin{equation}
\label{integrated_beta_function}
\ln\frac{\mu_2}{\mu_1} \simeq f\left(\alphas(\mu_2)\right)-f\left(\alphas(\mu_1)\right),
\end{equation}
with
\begin{equation}
\label{f_at_three_loops}
f(x)=-\frac{b_1}{2b_0^2} \ln \left( \frac{x^2}{b_0+b_1x+b_2x^2} \right) - \frac{1}{b_0x} + \frac{b_1^2-2b_0 b_2}{b_0^2\sqrt{-\Delta}} \arctan \left( \frac{b_1+2b_2x}{\sqrt{-\Delta}} \right),
\end{equation}
where, introducing $c_1= 0.94327(4)$ and $c_2= 0.5216(5)+c_1^2$, we defined
\begin{equation}
\label{three-loop_SU2_beta_function_coefficients}
b_0=-\frac{11}{3\pi},\quad b_1=-\frac{17}{3\pi^2} ,\quad b_2=-\left[ \frac{2857}{216\pi^3} +\frac{17}{3\pi^2}c_1 +\frac{11}{3\pi}(c_1^2-c_2) \right],\quad \mbox{and } \Delta=b_1^2-4 b_0 b_2.
\end{equation}
Our numerical results confirm the accuracy of the three-loop (and two-loop) perturbative $\beta$ function down to $\mu \sim 1$~GeV.

Similarly, the plot on the right-hand side of fig.~\ref{fig:results} shows our numerical results for $\alphas=g^2/(4\pi)$ in $\SU(3)$ Yang-Mills theory, and their comparison with the integrated version of the perturbative prediction~\cite{Bode:1998hd}:
\begin{equation}
\label{three-loop_SU3_SF_beta_function}
\frac{d \alphas}{d ( \ln \mu)} = -\frac{11}{2\pi}\alphas^2 -\frac{51}{4\pi^2}\alphas^3 -0.966(18)\alphas^4 +O ( \alphas^5 )
\end{equation}
and its truncations at $O(\alphas^3)$ and at $O(\alphas^2)$. Also in this case, our results are in complete agreement with those obtained from conventional Monte~Carlo calculations~\cite{Luscher:1993gh}, and confirm the validity of the three-loop perturbative prediction down to scales $\mu \sim 1$~GeV.

\section{Summary and concluding remarks}

Jarzynski's theorem provides a rigorous basis for non-equilibrium Monte~Carlo algorithms that can be used for various applications in lattice field theory. As an example, in this presentation we discussed the study reported in ref.~\cite{Francesconi:2020fgi}, in which Jarzynski's equality was used to compute the quantum effective action in the Schr\"odinger-functional formalism, from which one can determine the running coupling in this scheme. Our results for $\SU(2)$ and $\SU(3)$ Yang-Mills theories are fully compatible with those obtained from standard (equilibrium) Monte~Carlo calculations~\cite{Luscher:1992zx, Luscher:1993gh}. As detailed in ref.~\cite{Francesconi:2020fgi}, our non-equilibrium algorithm proves computationally very efficient, and we expect that it may reveal its full power especially in situations where standard Monte~Carlo calculations are hampered by particularly challenging computational complexity problems. Finally, while the present study was carried out in purely gluonic theories, the extension to full QCD is straightforward, and the generalization to a diverse array of other physical quantities is possible.

\acknowledgments
O.~F. is thankful for support by the Independent Research Fund Denmark, Research Project 1, grant number 8021-00122B.
\bibliography{paper}

\end{document}